\documentclass[final,5p,times,twocolumn,numbers,sort, compress]{elsarticle}
\usepackage{amssymb}
\usepackage{lipsum}
\usepackage{xcolor}
\usepackage{mathtools}
\usepackage{amsmath}
\usepackage[utf8]{inputenc}
\usepackage{subfig}
\usepackage{float}
\journal{Physics Letters A}
\begin{document}
\begin{frontmatter}

\title{A phenomenological description of critical slowing down at period-doubling bifurcations}
\author{$^1$Edson D.\ Leonel, $^1$Jo\~ao P. C. Ferreira, $^2$Diego F. M. Oliveira}
\address{{$^1$Departamento de Física, UNESP - Universidade Estadual Paulista, Avenida 24A, 1515, Bela Vista, Rio Claro, 13506-900, São Paulo, Brazil\\
$^2$School of Electrical Engineering and Computer Science - University of North Dakota, Grand Forks, Avenue Stop 8357, 58202, North Dakota, USA.}}

\begin{abstract}
We present a phenomenological description of the critical slowing down associated with period-doubling bifurcations in discrete dynamical systems. Starting from a local Taylor expansion around the fixed point and the bifurcation parameter, we derive a reduced description that captures the convergence towards stationary state both at and near criticality. At the bifurcation point, three universal critical exponents are obtained, characterising the short-time behaviour, the asymptotic decay, and the crossover between these regimes. Away from criticality, a fourth exponent governing the relaxation time is identified. We show this phenomenology, well established for one-dimensional maps, extends naturally to two-dimensional mappings. By projecting the dynamics onto the centre manifold, we demonstrate that the local normal form of a two-dimensional period-doubling bifurcation reduces to the same universal structure found in one dimension. The theoretical predictions are validated numerically using the H\'enon and Ikeda maps, showing excellent agreement for all scaling laws and critical exponents.
\end{abstract}

\end{frontmatter}

\section{Introduction}
\label{intro}

Bifurcations constitute fundamental mechanisms through which nonlinear
dynamical systems undergo qualitative changes in their behavior as control
parameters are varied \cite{kuznetsov1998elements,strogatz2024nonlinear}. Such transitions are
ubiquitous in nature and arise across a wide range of contexts, including
biological systems \cite{tyson2001regulation}, chemical reactions far from
equilibrium \cite{prigogine1978time}, engineering applications involving
stability and buckling phenomena \cite{thompson1963basic}, and
socio-economic models governed by nonlinear feedback mechanisms
\cite{lorenz1993nonlinear}.
In biology, bifurcations have been identified in population dynamics, neural activity, and epidemiological models, where they often signal transitions between stable and oscillatory regimes or the onset of complex temporal behaviour~\cite{murray2002,strogatz2024nonlinear}. In chemistry, bifurcations play a central role in autocatalytic reactions and chemical oscillators, such as the Belousov--Zhabotinsky reaction, governing the emergence of periodic and chaotic dynamics~\cite{epstein1998}. Engineering systems frequently exhibit bifurcations in feedback-controlled devices, power grids, and mechanical oscillators, where the loss of stability may lead to undesirable or even catastrophic behaviour~\cite{nayfeh2008}. Bifurcation phenomena have also been reported in economic and social systems,
in which parameter variations can trigger abrupt transitions between distinct
macroscopic states~\cite{stachurski2009, Gandolfo1996EconomicDynamics,BrockHommes1998, Granovetter1978,Durlauf2001,Scheffer2009}.

From the perspective of dynamical systems theory, bifurcations are commonly classified into two broad categories: local and global bifurcations. Local bifurcations are characterised by qualitative changes that are fully determined by the behaviour of the system in an arbitrarily small neighbourhood of an invariant object, such as a fixed point or a periodic orbit. These bifurcations can be detected and analysed through local stability analysis, typically based on the linearisation of the dynamics and the
spectral properties of the associated Jacobian matrix~\cite{Hirsch2013,Hale1991,Perko2001}. In contrast, global bifurcations involve large-scale reorganisations of phase space and depend on the global structure of invariant manifolds. Examples include homoclinic and heteroclinic bifurcations, which cannot be inferred from local information alone and often give rise to complex dynamical behaviour~\cite{wiggins2003}.

A particularly important dynamical feature associated with bifurcations
is the phenomenon of critical slowing down~\cite{Hohenberg1977,
Goldenfeld1992,Kuehn2011}. As a system approaches a bifurcation point, its recovery rate from small perturbations decreases, leading to a divergence of characteristic relaxation times~\cite{strogatz2024nonlinear}. Beyond serving as a clear dynamical precursor of a bifurcation, critical slowing down also plays a central role in the emergence of universal scaling laws near criticality.

Within the framework of discrete dynamical systems, one-dimensional maps have long served as paradigmatic models for understanding bifurcation phenomena and routes to chaos. In particular, the period-doubling cascade and its associated universality properties, first identified in unimodal maps, represent a cornerstone of nonlinear dynamics~\cite{feigenbaum1978,collet1980}. In the vicinity of a period-doubling bifurcation, the dynamics of one-dimensional maps can be described by a universal normal form, reflecting the fact that systems with distinct microscopic details may nevertheless exhibit identical scaling behaviour close to criticality.

The purpose of the present work is to extend this concept of universality to higher-dimensional discrete systems. Specifically, we investigate the normal form associated with period-doubling bifurcations in two-dimensional maps. We demonstrate that the universal features observed in one-dimensional systems persist in higher dimensions. Despite the increased dimensionality and the richer phase-space structure inherent to two-dimensional maps, we show that the local dynamics near a period-doubling bifurcation can still be reduced to the same universal form that governs one-dimensional maps. This result reinforces the robustness of universality in bifurcation theory and provides further insight into the local mechanisms underlying complex dynamics in multidimensional discrete systems.

This paper is organised as follows. Section~\ref{sec2} presents the analytical procedure based on a Taylor expansion performed around both the fixed point and the bifurcation parameter. Two complementary descriptions are considered: (i) the dynamics exactly at the bifurcation, leading to a set of three universal critical exponents that characterise the convergence to the stationary state, and (ii) the dynamics in the vicinity of the bifurcation, which allows the determination of a fourth critical exponent. Together, these four exponents define the universality class of the bifurcation. Section~\ref{disc} discusses a representative example using the H\'enon and the Ikeda maps. Conclusions and final remarks are presented in Section~\ref{conc}.

\section{A phenomenological description of the critical exponents}
\label{sec2}

In this section, we present the procedure used to characterise the convergence towards the stationary state both at and in the vicinity of bifurcation points. We begin with the analysis of one-dimensional mappings and subsequently extend the approach to two-dimensional systems. For clarity, we first focus on the simplest and most general case.

\subsection{Local analysis}

Let the dynamics be described by a one-dimensional map of the form
\begin{equation}
x_{n+1}=f(x_n,r),
\label{eq1}
\end{equation}
where $f$ is a nonlinear function of the dynamical variable $x$ and the control parameter $r$. We assume that, at $r=r_c$, a local bifurcation occurs at a fixed point defined by $f(x^*,r_c)=x^*$. A local bifurcation corresponds to a qualitative change in the system dynamics induced by variations of a control parameter, leading to a change in the stability of an invariant object, such as a fixed point or a periodic orbit, while the analysis remains restricted to an arbitrarily small neighbourhood of that object.

The central idea is to perform a Taylor expansion of Eq.~(\ref{eq1}) around both the fixed point $x^*$ and the critical parameter value $r_c$, and to analyse the resulting local dynamics. To this end, we write $x_n=x^*+\epsilon_n$ and $r=r_c+\mu$, where both $\epsilon_n$ and $\mu$ are assumed to be sufficiently small. Substituting these expressions into Eq.~(\ref{eq1}) and expanding up to third order in $\epsilon_n$ and first order in $\mu$, we obtain
\begin{eqnarray}
x^*&+&\epsilon_{n+1}=f(x^*,r_c)+\left.\epsilon_n{{df}\over{dx}}\right|_{*}+\left.\mu{{df}\over{dr}}\right|_{*}+\left.{{\epsilon_n^2}\over{2}}{{d^2f}\over{dx^2}}\right|_{*}+\nonumber\\
&+&\left.2\epsilon_n\mu{{d^2f}\over{dxdr}}\right|_{*}+\left.{{\epsilon_n^3}\over{6}}{{d^3f}\over{dx^3}}\right|_{*}+\ldots ,
\label{eq2}
\end{eqnarray}
where the symbol ($^*$) indicates that all derivatives are evaluated at $(x^*,r_c)$. Higher-order contributions involving terms such as $\mu^2$, $\mu^2\epsilon_n$, $\mu\epsilon_n^2$, and higher powers have been neglected.

At a period-doubling bifurcation, the linear term satisfies $\left.{{df}\over{dx}}\right|_{*}=-1$, and the map can therefore be written as
\begin{equation}
\epsilon_{n+1}=-\epsilon_n+j_2\mu+j_3\epsilon_n^2+j_4\epsilon_n\mu+j_5\epsilon_n^3,
\label{eq3}
\end{equation}
where $j_1=\left.{{df}\over{dx}}\right|_{*}=-1$, $j_2=\left.{{df}\over{dr}}\right|_{*}$, $j_3={{1}\over{2}}\left.{{d^2f}\over{dx^2}}\right|_{*}$, $j_4=2\left.{{d^2f}\over{dxdr}}\right|_{*}$, and $j_5={{1}\over{6}}\left.{{d^3f}\over{dx^3}}\right|_{*}$.

The convergence to the stationary state at the period-doubling bifurcation is analysed through the second iterate of the map, namely $x_{n+2}=f(x_{n+1},r)$. Applying this iteration to Eq.~(\ref{eq3}) yields
\begin{equation}
\epsilon_{n+2}=\epsilon_n-2\mu\epsilon_n(j_2j_3+j_4)-\epsilon_n^3(1+2j_3^2+j_5),
\label{eq4}
\end{equation}
where terms of order higher than $\epsilon_n^3$ have been neglected.

We now distinguish two cases: (i) the dynamics exactly at the bifurcation point, corresponding to $\mu=0$, and (ii) the dynamics in the vicinity of the bifurcation, where $\mu<0$ or $\mu>0$. We begin by analysing the first case.

\subsection{Convergence to the stationary state at the bifurcation}

The case $\mu=0$ corresponds to the dynamics precisely at the local bifurcation, which in the present context is a period-doubling bifurcation. For notational convenience, we define the coefficients $A=2(j_2j_3+j_4)$ and $B=1+2j_3^2+j_5$. Furthermore, we assume that, sufficiently close to the stationary state, the dynamics evolves slowly, allowing the discrete difference equation to be approximated by a differential equation,
\begin{equation}
\epsilon_{n+2}-\epsilon_n\cong{{d\epsilon}\over{dn}}=-B\epsilon^3.
\label{eq5}
\end{equation}

Integrating both sides of Eq.~(\ref{eq5}) yields
\begin{equation}
\int_{\epsilon_0}^{\epsilon(n)}{{d\epsilon}\over{\epsilon^3}}=-B\int_0^n dn^{\prime},
\label{eq6}
\end{equation}
from which, after straightforward algebraic manipulation, we obtain
\begin{equation}
\epsilon(n)={{\epsilon_0}\over{\sqrt{1+2B\epsilon_0^2n}}},
\label{eq7}
\end{equation}
valid for $B>0$.

Equation~(\ref{eq7}) allows the identification of three distinct dynamical regimes. The first regime corresponds to the condition $2B\epsilon_0^2n\ll1$, for which $\epsilon(n)\propto\epsilon_0^{\alpha}$. In this regime, the exponent $\alpha$ characterises the short-time behaviour of the distance from the equilibrium state. Since the expression is independent of $n$, one finds $\alpha=1$, in agreement with previous studies~\cite{r1,teixeira2015convergence}.

The second regime is observed when $2B\epsilon_0^2n\gg1$. In this limit, Eq.~(\ref{eq7}) reduces to $\epsilon(n)\cong{{1}\over{\sqrt{2B}}}n^{\beta}$, where $\beta$ governs the rate at which the distance to the stationary state decays. For the period-doubling bifurcation, the exponent $\beta=-1/2$ is universal~\cite{r1}.

The third regime corresponds to the crossover between the initial plateau and the asymptotic decay, occurring at an iteration number $n_x={{1}\over{2B}}\epsilon_0^{z}$. The exponent $z$ defines the crossover scaling and is also universal, taking the value $z=-2$ for the period-doubling bifurcation~\cite{r1}.

The universality of these three critical exponents was previously discussed in Ref.~\cite{r1} for a logistic-like map of the form $x_{n+1}=Rx_n(1-x_n^{\gamma})$, with $\gamma\ge1$. The aim of the present work is to demonstrate that this phenomenology is more general and extends naturally to two-dimensional mappings. To illustrate the applicability of the approach in higher dimensions, we consider the H\'enon~\cite{r3} map, with the corresponding discussion presented in Section~\ref{disc}.

\subsection{Convergence to the steady state close to the bifurcation}

In the previous section, we analysed the convergence to the steady state exactly at the bifurcation point. We now turn our attention to the convergence in the vicinity of the bifurcation, which may occur either before or after the critical parameter value. For $\mu\neq0$, the effective map can be written as
\begin{equation}
\epsilon_{n+2}=\epsilon_n-\mu\epsilon_nA-\epsilon_n^3B=h(\epsilon_n).
\label{eq8}
\end{equation}

The fixed points of Eq.~(\ref{eq8}) are obtained by imposing $\epsilon_{n+2}=\epsilon_n$. This condition yields three fixed points: $\epsilon_1=0$ and $\epsilon_{2,3}=\pm\sqrt{-{{\mu A}\over{B}}}$, which exist provided that $\mu A<0$. The fixed point $\epsilon_1$ is asymptotically stable when $0<\mu A<2$, a condition that is satisfied either if (i) $\mu<0$ and $A<0$, or if (ii) $\mu>0$ and $A>0$. 

On the other hand, the fixed points $\epsilon_{2,3}$ exist only when either (i) $\mu>0$ and $A<0$, or (ii) $\mu<0$ and $A>0$. These fixed points are asymptotically stable if $-{{1}\over{3}}<\mu A<0$. An important consequence of these existence and stability conditions is that the amplitude of the stable fixed points scales as $\epsilon\sim\sqrt{|\mu|}$, a behaviour that can be directly measured numerically (see, for instance, Fig.~\ref{Fig6}).

We now derive an analytical expression for $\epsilon(n)$ by transforming the discrete difference equation into an ordinary differential equation, following the same procedure adopted in the previous section. Assuming once again that, sufficiently close to the fixed point, the orbit converges slowly from $n$ to $n+1$, we approximate
\begin{equation}
\epsilon_{n+2}-\epsilon_n\cong{{d\epsilon}\over{dn}}=-\mu A\epsilon-B\epsilon^3.
\label{eq9}
\end{equation}
Equation~(\ref{eq9}) is a first-order differential equation and can be readily integrated under appropriate initial conditions. Performing the integration yields
\begin{equation}
\int_{\epsilon_0}^{\epsilon(n)}{{d\epsilon}\over{\epsilon(\mu A+B\epsilon^2)}}=\int_0^n dn^{\prime},
\label{eq10}
\end{equation}
from which, after straightforward algebraic manipulation, we obtain
\begin{equation}
\epsilon(n)={{\epsilon_0 e^{-A\mu n}}\over{\sqrt{1+{{B\epsilon_0^2}\over{A\mu}}\left(1-e^{-2A\mu n}\right)}}}.
\label{eq11}
\end{equation}

We now discuss the range of validity of Eq.~(\ref{eq11}). For the fixed point $\epsilon_1$, corresponding to the regime prior to the period-doubling bifurcation, the condition $0<\mu A<2$ implies $\mu A>0$, and therefore the exponential term in the numerator decays to zero as $n$ increases.

It is well known~\cite{r1} that, close to a bifurcation, the distance from the stationary state typically behaves as
\begin{equation}
d(n)\cong d_0 e^{-n/\tau},
\label{eq12}
\end{equation}
where the characteristic relaxation time scales as $\tau\propto\mu^{\delta}$, with $\delta$ denoting a critical exponent. By comparing Eqs.~(\ref{eq12}) and (\ref{eq11}), we conclude that $\delta=-1$ for the period-doubling bifurcation. Notably, this result does not depend on the specific functional form of the mapping, indicating that the exponent $\delta$ is universal for the period-doubling bifurcation.

After the bifurcation, the stable fixed points are $\epsilon_{2,3}$. In this regime, the argument of the exponential in Eq.~(\ref{eq11}) becomes negative, since asymptotic stability requires $-{{1}\over{3}}<\mu A<0$. As a consequence, both the numerator and the denominator of Eq.~(\ref{eq11}) diverge as $n$ becomes sufficiently large. To circumvent this difficulty, the expression can be conveniently rewritten as
\begin{equation}
\epsilon^2(n)=-{{\mu A}\over{B}}{{1}\over{1-e^{2\mu An}\left({{\mu A}\over{B\epsilon_0^2}}+1\right)}},
\label{eq13}
\end{equation}
which has the same mathematical structure as $f(x)=1/(1-x)=(1-x)^{-1}\cong(1+x)$ for small $x$. Applying this expansion to Eq.~(\ref{eq13}) leads to
\begin{equation}
\epsilon^2(n)+{{\mu A}\over{B}}=-{{\mu A}\over{B}}\left(1-{{\mu A}\over{B\epsilon_0^2}}\right)e^{2A\mu n},
\label{eq14}
\end{equation}
valid for $-{{1}\over{3}}<\mu A<0$.

Equation~(\ref{eq14}) has the same mathematical form as Eq.~(\ref{eq12}), allowing us to conclude once again that the critical exponent governing the relaxation time is $\delta=-1$.

\subsection{Period-doubling bifurcation in two-dimensional maps}

We now discuss a natural generalisation of the previous analysis to two-dimensional mappings. We consider a map of the form
\begin{equation}
\mathbf{z}_{n+1} = \mathbf{F}(\mathbf{z}_n,r)~,~
\mathbf{z}=(x,y)^T,
\end{equation}
where $\mathbf{F}=(F_1,F_2)^T$, with $F_1$ and $F_2$ being nonlinear functions of the variables $x$ and $y$ and of the control parameter $r$. The superscript $T$ denotes the transpose of the vector. We assume the existence of a fixed point $\mathbf{z}^*$ satisfying
\begin{equation}
\mathbf{F}(\mathbf{z}^*,r_c)=\mathbf{z}^*.
\end{equation}

Defining $\mathbf{u}_n=\mathbf{z}_n-\mathbf{z}^*$, a Taylor expansion of the map around $(\mathbf{z}^*,r_c)$ yields
\begin{equation}
\mathbf{u}_{n+1}
= \tilde{A}\mathbf{u}_n
+ \frac{1}{2}\tilde{B}(\mathbf{u}_n,\mathbf{u}_n)
+ \frac{1}{6}C(\mathbf{u}_n,\mathbf{u}_n,\mathbf{u}_n)
+ \mathcal{O}(\|\mathbf{u}_n\|^4),
\label{eq:2d_taylor}
\end{equation}
where
\begin{equation}
\tilde{A} = D_{\mathbf{z}}\mathbf{F}(\mathbf{z}^*,r_c)
\end{equation}
is the Jacobian matrix, while $\tilde{B}$ and $C$ denote the symmetric bilinear and trilinear forms constructed from the second and third derivatives of $\mathbf{F}$, respectively.

For each $i\in\{1,2\}$, let $H_i$ denote the Hessian matrix of $F_i$ with respect to $(x,y)$,
\begin{equation}
H_i
=
D^2_{\mathbf{z}}F_i(\mathbf{z}^*,r_c)
=
\begin{pmatrix}
F_{i,xx} & F_{i,xy}\\
F_{i,yx} & F_{i,yy}
\end{pmatrix}_{(\mathbf{z}^*,r_c)}.
\end{equation}
Then, for arbitrary vectors $\mathbf{u},\mathbf{v}\in\mathbb{R}^2$,
\begin{equation}
\tilde{B}(\mathbf{u},\mathbf{v})
=
\begin{pmatrix}
\mathbf{u}^T H_1 \mathbf{v}\\[2mm]
\mathbf{u}^T H_2 \mathbf{v}
\end{pmatrix},
\label{eq:B_def}
\end{equation}
which is symmetric because each $H_i$ is symmetric ($F_{i,xy}=F_{i,yx}$ for $C^2$ maps).

In component form,
\begin{align}
B_1(\mathbf{u},\mathbf{v})
&=
F_{1,xx}\,u_1v_1
+F_{1,xy}\,(u_1v_2+u_2v_1)
+F_{1,yy}\,u_2v_2,
\\
B_2(\mathbf{u},\mathbf{v})
&=
F_{2,xx}\,u_1v_1
+F_{2,xy}\,(u_1v_2+u_2v_1)
+F_{2,yy}\,u_2v_2,
\end{align}
with all derivatives evaluated at $(\mathbf{z}^*,r_c)$.

Let $T_i$ denote the third-derivative tensor of $F_i$ with respect to $(x,y)$, whose components are defined as
\begin{equation}
(T_i)_{jkl}
=
\frac{\partial^3 F_i}{\partial z_j \partial z_k \partial z_l}(\mathbf{z}^*,r_c),
\qquad
z_1=x,\ z_2=y,
\qquad
j,k,l\in\{1,2\}.
\end{equation}
Then, for $\mathbf{u},\mathbf{v},\mathbf{w}\in\mathbb{R}^2$,
\begin{equation}
C(\mathbf{u},\mathbf{v},\mathbf{w})
=
\begin{pmatrix}
\displaystyle \sum_{j,k,l=1}^2 (T_1)_{jkl}\,u_j v_k w_l\\[3mm]
\displaystyle \sum_{j,k,l=1}^2 (T_2)_{jkl}\,u_j v_k w_l
\end{pmatrix}.
\label{eq:C_def}
\end{equation}
This form is symmetric in its arguments when $F_i\in C^3$, as a consequence of the equality of mixed partial derivatives. Here $\mathbf{u}$, $\mathbf{v}$ and $\mathbf{w}$ denote arbitrary vectors in $\mathbb{R}^2$ associated with infinitesimal displacements in phase space. The vector $\mathbf{w}$ is introduced solely to make explicit the trilinear character of $C$; in the actual Taylor expansion one evaluates $C(\mathbf{u},\mathbf{u},\mathbf{u})$, with the parameter fixed at $r=r_c$.

Equivalently, writing explicitly the independent third derivatives, one obtains
\begin{align}
C_i(\mathbf{u},\mathbf{v},\mathbf{w})
&=
F_{i,xxx}\,u_1 v_1 w_1
+F_{i,xxy}\,\big(u_1 v_1 w_2 + u_1 v_2 w_1 + u_2 v_1 w_1\big)
\nonumber\\
&\quad
+F_{i,xyy}\,\big(u_1 v_2 w_2 + u_2 v_1 w_2 + u_2 v_2 w_1\big)
+F_{i,yyy}\,u_2 v_2 w_2,
\label{eq:C_explicit}
\end{align}
for $i=1,2$, with all derivatives evaluated at $(\mathbf{z}^*,r_c)$.

At a period-doubling bifurcation, the Jacobian matrix has eigenvalues
\begin{equation}
\lambda_1=-1, \qquad |\lambda_2|<1,
\end{equation}
implying that the centre manifold is one-dimensional. Let $\mathbf{q}$ and $\mathbf{p}$ denote the right and left eigenvectors associated with $\lambda_1=-1$, normalised such that
\begin{equation}
\mathbf{p}^T\mathbf{q}=1.
\end{equation}

When the dynamics is projected onto the centre manifold, the state vector can be expressed as
\begin{equation}
\mathbf{u} = s\,\mathbf{q} + h(s,\mu),
\qquad \mu = r-r_c,
\end{equation}
where $h(s,\mu)=\mathcal{O}(s^2,\mu s)$ accounts for the slaving of the stable direction.

Substituting this expression into Eq.~(\ref{eq:2d_taylor}) and projecting along $\mathbf{p}$ yields the reduced one-dimensional map
\begin{equation}
s_{n+1}
= -s_n + \tilde{\alpha} \mu s_n + \tilde{a} s_n^2 + \tilde{b} s_n^3
+ \mathcal{O}(s_n^4,\mu s_n^2),
\label{eq:center_reduction}
\end{equation}
where the coefficients $\tilde{a}$ and $\tilde{b}$ are explicit combinations of the tensors $\tilde{B}$ and $C$, together with contributions induced by the stable eigenmode.

Setting $\mu=0$ and iterating Eq.~(\ref{eq:center_reduction}) twice leads to
\begin{equation}
s_{n+2}
= s_n + \tilde{\gamma} s_n^3 + \mathcal{O}(s_n^4),
\end{equation}
with
\begin{equation}
\tilde{\gamma} = -2(\tilde{a}^2+\tilde{b}).
\end{equation}

Thus, as in the one-dimensional case, the quadratic term cancels identically in the second iterate, even though $\tilde{a}\neq 0$ in general. The period-doubling bifurcation is therefore governed by the cubic term of the reduced dynamics.

Considering the dynamics near the bifurcation, the normal form of the second iterate becomes
\begin{equation}
s_{n+2}
= s_n + \kappa \mu s_n + \tilde{\gamma} s_n^3 + \cdots,
\end{equation}
which fully characterises the local structure of the period--2 orbit.

\section{Discussions}
\label{disc}

This section is devoted to the discussion of the results obtained in the previous sections and to their extension to two-dimensional mappings. Our numerical investigations are carried out using the H\'enon map~\cite{r3} as a representative example.

\subsection{The H\'enon map}

The H\'enon map is defined as
\begin{equation}
\begin{cases}
x_{n+1} = 1 - a x_n^2 + y_n, \\
y_{n+1} = b x_n,
\end{cases}
\label{eq15}
\end{equation}
where $a$ and $b$ are control parameters. The Jacobian matrix of the map is given by
\begin{equation}
J =
\begin{pmatrix}
-2 a x_n & 1 \\
b & 0
\end{pmatrix}.
\end{equation}
The map preserves area only when $b=-1$. The parameter value originally considered by H\'enon~\cite{r3} is $b=0.3$, which we adopt throughout this work.

\begin{figure}[t]
\centerline{\includegraphics[width=1.0\linewidth]{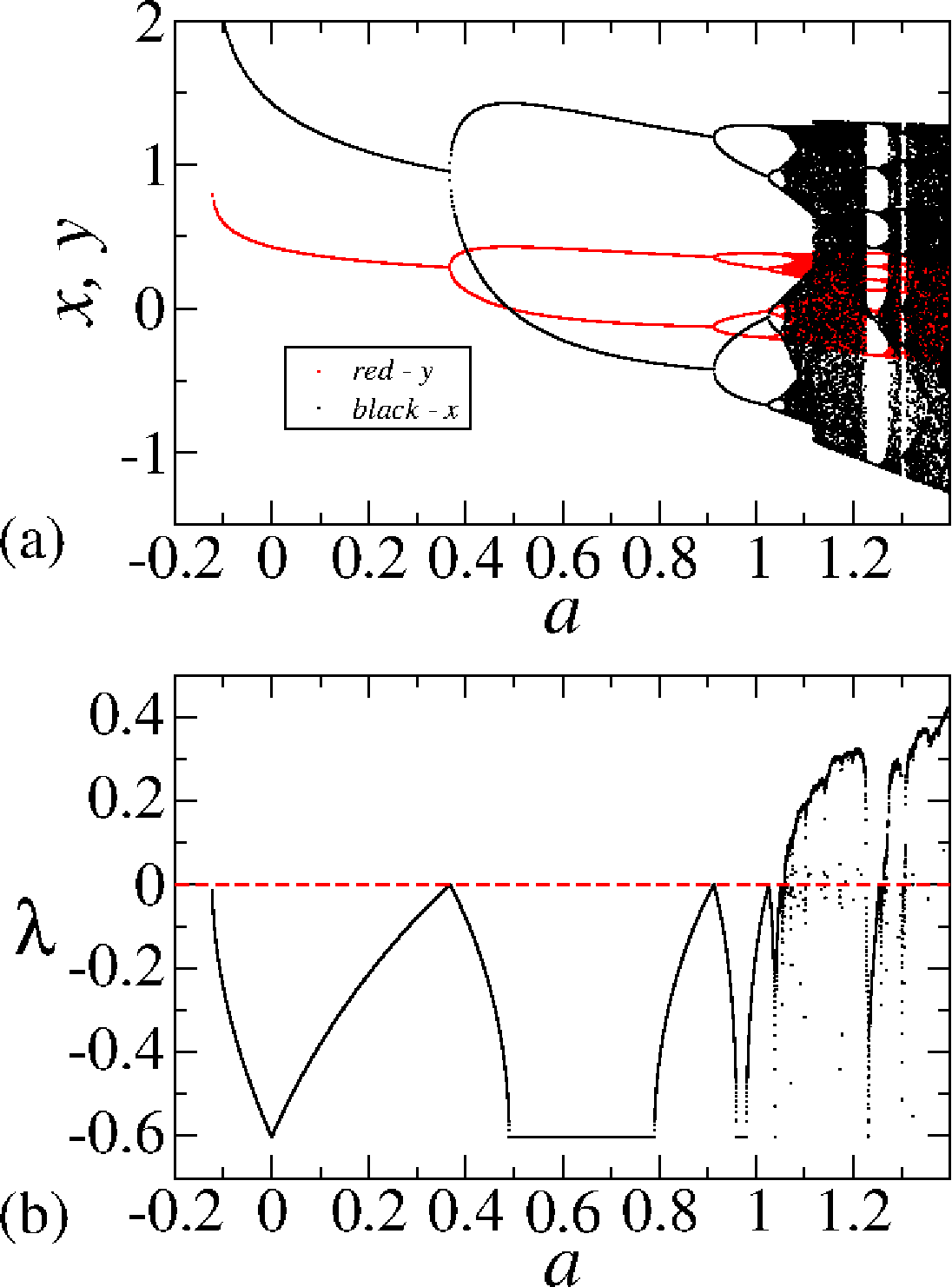}}
\caption{(a) Orbit diagram for the H\'enon map~(\ref{eq15}). (b) Largest Lyapunov exponent corresponding to the orbits shown in panel (a). The initial condition used was $(x_0,y_0)=(0.952723,0.2857)$ for a fixed $b=0.3$ and $a\in[-0.1,1.4]$. Each initial condition was iterated for $10^9$ iterations and only the last $100$ points of the orbit were recorded.}
\label{Fig1}
\end{figure}

Figure~\ref{Fig1}(a) shows the orbit diagram for both variables $x$ and $y$ as a function of the control parameter $a$.

\begin{figure}[t]
\centerline{\includegraphics[width=1.0\linewidth]{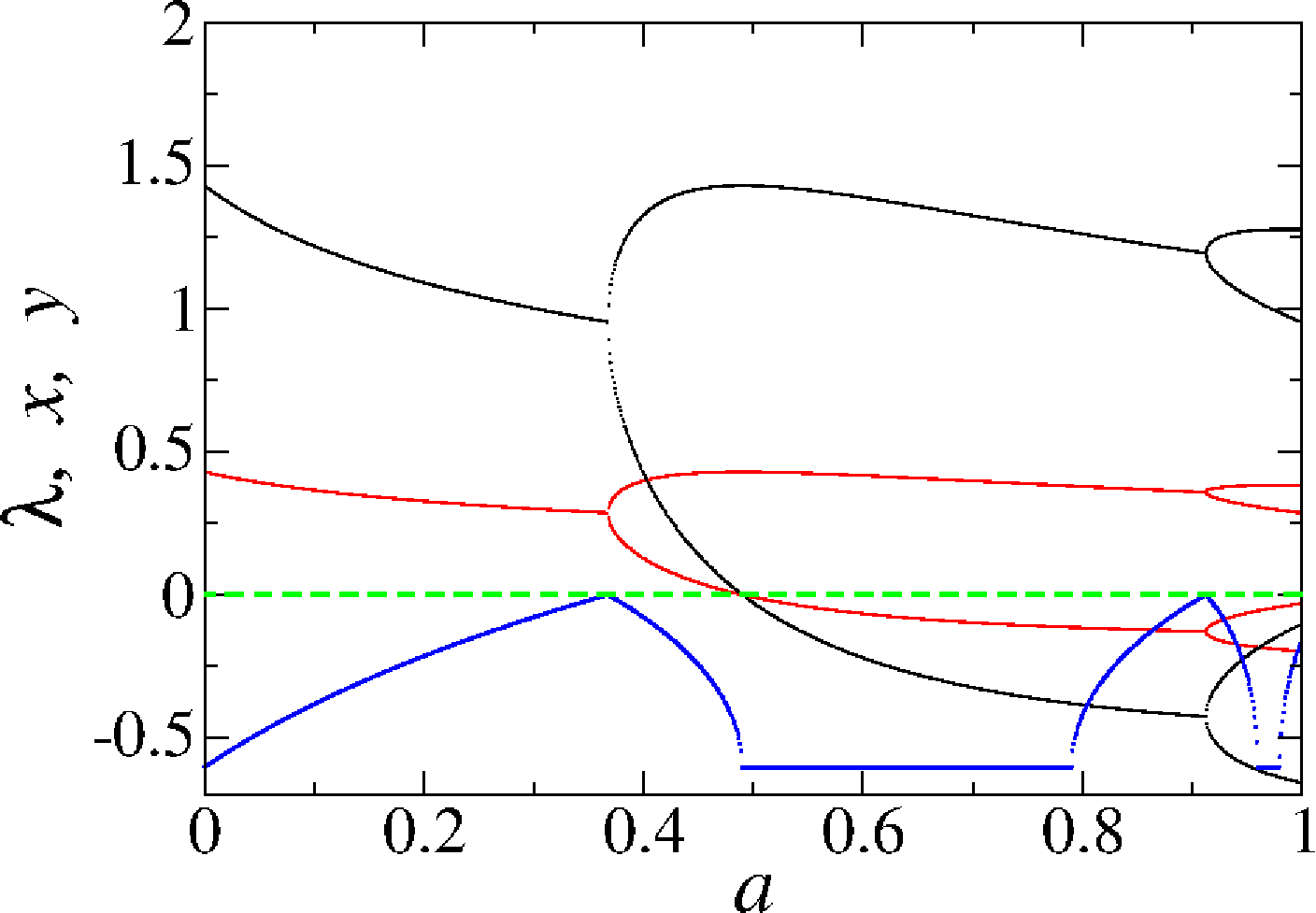}}
\caption{Orbit diagram for the H\'enon map~(\ref{eq15}) -- a zoom of Fig.~\ref{Fig1}(a) -- together with the largest Lyapunov exponent near the first period-doubling bifurcation. The dashed line is shown as a guide to the eye. The first period-doubling bifurcation is observed at $a=0.367495$.}
\label{Fig2}
\end{figure}

From Fig.~\ref{Fig1}(a), it is clear that both variables $x$ and $y$ undergo bifurcations at the same parameter values. At the bifurcation points, the largest Lyapunov exponent vanishes, reflecting the fact that one eigenvalue of the Jacobian matrix is equal to $\pm1$. The region of interest for the present study lies in the vicinity of the first period-doubling bifurcation. Figure~\ref{Fig2} displays a magnified view of Fig.~\ref{Fig1}(a), together with the corresponding largest Lyapunov exponent.

\begin{figure}[t]
\centerline{(a)\includegraphics[width=1.0\linewidth]{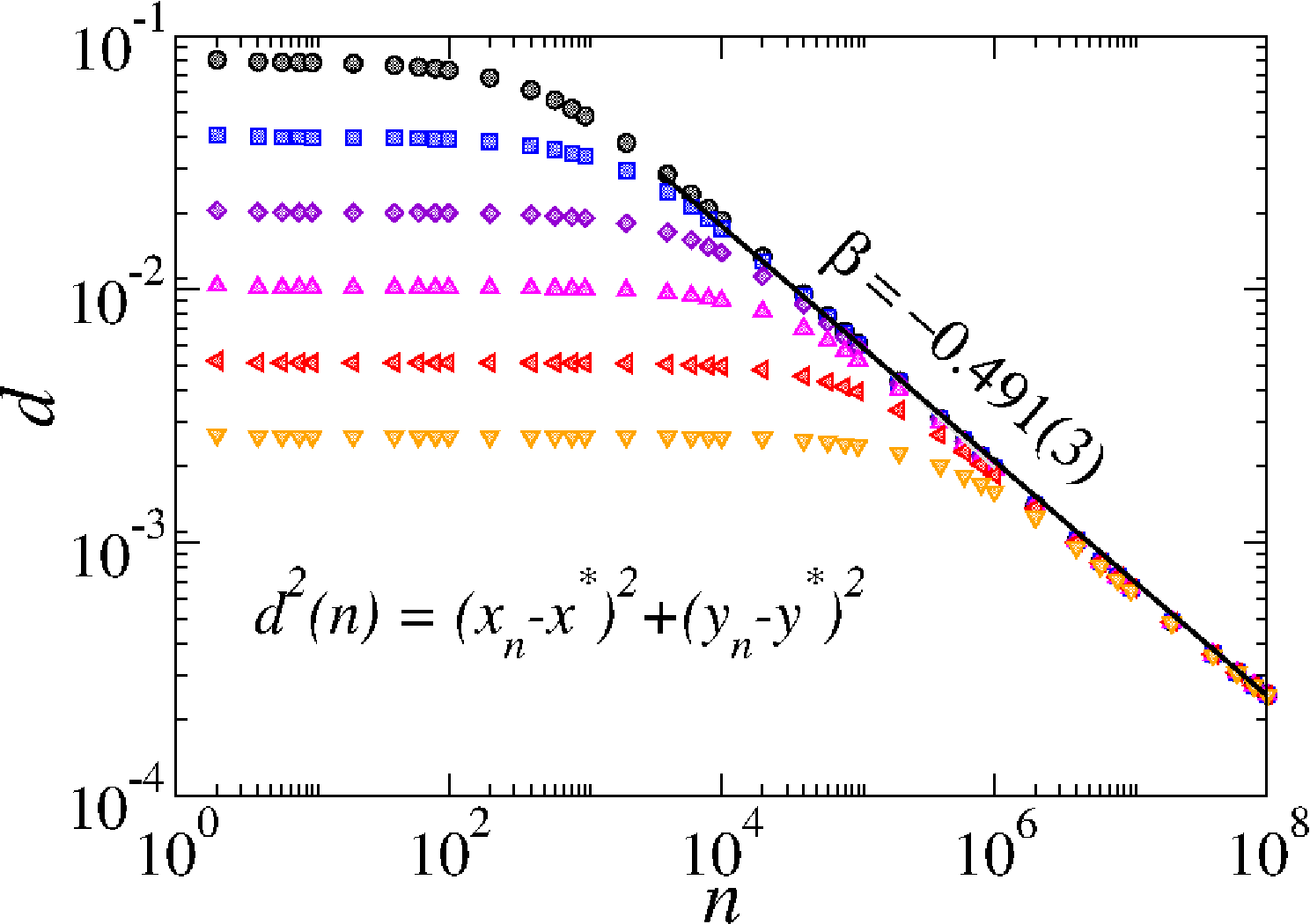}}
\centerline{(b)\includegraphics[width=1.0\linewidth]{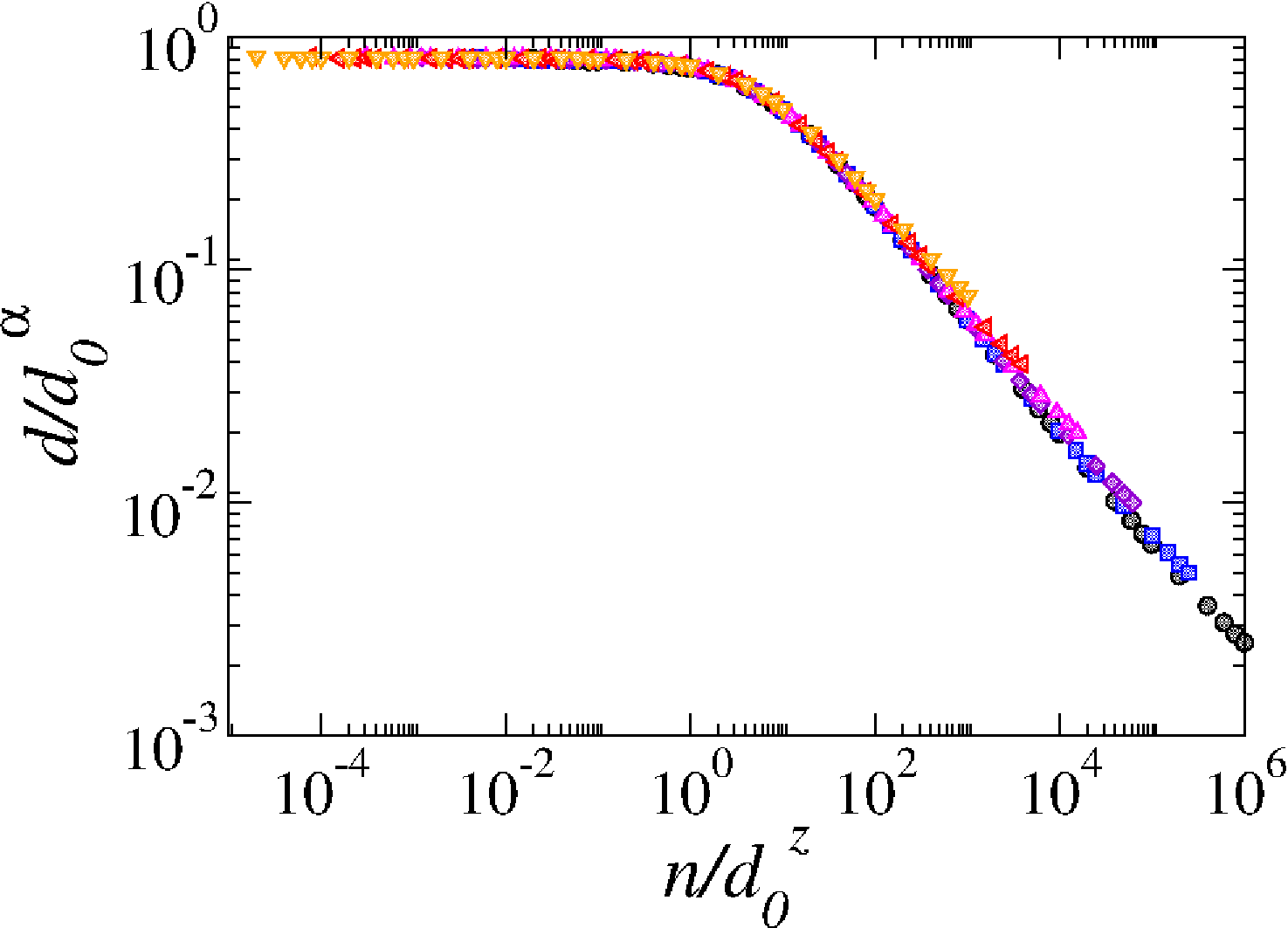}}
\caption{(a) Distance $d(n)$ as a function of the iteration number $n$ for different initial conditions near the fixed point at the period-doubling bifurcation of the H\'enon map. (b) Collapse of the curves shown in panel (a) onto a universal plot after the appropriate scaling transformations. The control parameters used were $b=0.3$ and $a=0.367495$.}
\label{Fig3}
\end{figure}

The Lyapunov exponents \cite{eckmann1985ergodic} are computed from
\begin{equation}
\lambda_j = \lim_{n \rightarrow \infty} \frac{1}{n} \ln |\Lambda_n^{(j)}|,
\label{c10_eq20}
\end{equation}
with $j = 1, 2$, where $\Lambda_n^{(j)}$ are the eigenvalues of the matrix
$M = \Pi_{i=1}^n J_i(x_i, y_i) = J_n J_{n-1} J_{n-2} \ldots J_2 J_1$.
Since the limit involves large values of $n$, the direct multiplication of the Jacobian matrices $J_i$ may result in numerical overflow or underflow, rendering the direct evaluation of $\lambda_j$ unreliable. To overcome this difficulty, we employ a triangularization procedure based on an orthogonal decomposition of the Jacobian matrix.

At each iteration, the Jacobian $J$ is factorised as $J = \Theta T$, where $\Theta$ is an orthogonal matrix and $T$ is an upper triangular matrix. The orthogonal matrix is written as
$$
\Theta = \left(\begin{array}{ll}
\cos(\theta)  &  -\sin(\theta)  \\
\sin(\theta)  &  \cos(\theta)\\
\end{array}\right),
$$
while the triangular matrix has the form
$$
T = \left(\begin{array}{ll}
T_{11}  &  T_{12}  \\
0       &  T_{22}\\
\end{array}\right).
$$

Using this decomposition, the product matrix $M$ can be rewritten as
\begin{eqnarray}
M &=& J_n J_{n-1} J_{n-2} \ldots J_2 J_1 \nonumber\\
  &=& J_n J_{n-1} J_{n-2} \ldots J_2 \Theta_1 \Theta_1^{-1} J_1,
\end{eqnarray}
which allows the definition $T_1 = \Theta_1^{-1} J_1$ and $\tilde{J}_2 = J_2 \Theta_1$. The elements of the matrix $T_1$ are obtained from
$$
\left(\begin{array}{ll}
T_{11}  &  T_{12}  \\
0       &  T_{22}\\
\end{array}\right)
=
\left(\begin{array}{ll}
\cos(\theta)  &  \sin(\theta)  \\
-\sin(\theta) &  \cos(\theta)\\
\end{array}\right)
\left(\begin{array}{ll}
j_{11}  &  j_{12}  \\
j_{21}  &  j_{22}\\
\end{array}\right).
$$

Imposing the condition $T_{21} = 0$ yields
$0 = -j_{11} \sin(\theta) + j_{21} \cos(\theta)$, which leads to
\begin{equation}
\frac{j_{21}}{j_{11}} = \frac{\sin(\theta)}{\cos(\theta)}.
\label{c10_eq21}
\end{equation}
Rather than computing $\theta = \arctan(j_{21}/j_{11})$, which is numerically inefficient, the trigonometric functions can be obtained directly from the Jacobian elements as
\begin{equation}
\cos(\theta) = \frac{j_{11}}{\sqrt{j_{11}^2 + j_{21}^2}}~~~,~~~
\sin(\theta) = \frac{j_{21}}{\sqrt{j_{11}^2 + j_{21}^2}}.
\end{equation}

With these expressions, the diagonal elements of the triangular matrix are given by
$T_{11} = j_{11} \cos(\theta) + j_{21} \sin(\theta)$ and
$T_{22} = -j_{12} \sin(\theta) + j_{22} \cos(\theta)$, yielding
\begin{equation}
T_{11} = \frac{j_{11}^2 + j_{21}^2}{\sqrt{j_{11}^2 + j_{21}^2}}~~~,~~~
T_{22} = \frac{j_{11} j_{22} - j_{12} j_{21}}{\sqrt{j_{11}^2 + j_{21}^2}}.
\end{equation}

Once the elements $T_{11}$ and $T_{22}$ are determined, the transformed Jacobian for the next iteration is computed as $\tilde{J}_2 = J_2 \Theta_1$, namely
$$
\left(\begin{array}{ll}
\tilde{j}_{11}  &  \tilde{j}_{12}  \\
\tilde{j}_{21}  &  \tilde{j}_{22}\\
\end{array}\right)
=
\left(\begin{array}{ll}
j_{11}  &  j_{12}  \\
j_{21}  &  j_{22}\\
\end{array}\right)
\left(\begin{array}{ll}
\cos(\theta)  &  -\sin(\theta)  \\
\sin(\theta)  &   \cos(\theta)\\
\end{array}\right).
$$

This procedure is iterated sequentially along the orbit, processing the entire sequence of Jacobian matrices $J_1, J_2, \ldots, J_n$. After all matrices have been triangularised, the Lyapunov exponents are finally obtained from
\begin{equation}
\lambda_j = \lim_{n \rightarrow \infty}{1\over{n}} \sum_{i=1}^{n} \ln |T_{jj}^{(i)}|, \quad j = 1, 2.
\end{equation}

The first period-doubling bifurcation of the H\'enon map occurs at the control parameter value $a=0.36749$, with the corresponding fixed point given by $x^*=0.95243773543628962$ and $y^*=0.28569725028983156$. Starting from an initial condition sufficiently close to this fixed point, and provided that it is asymptotically stable, the orbit converges towards it as time evolves. The distance from the fixed point is defined as
\begin{equation}
d(n)=\sqrt{(x_n-x^*)^2+(y_n-y^*)^2},
\label{EDist}
\end{equation}
and must decay to zero as the orbit approaches the steady state. Figure~\ref{Fig3} illustrates the convergence towards the fixed point at the bifurcation.

From Fig.~\ref{Fig3}(a), several characteristic features can be identified. For small iteration numbers $n$, the distance from the fixed point remains approximately constant up to a crossover iteration $n_x$, after which the curve bends towards a decaying regime as the orbit approaches the steady state. For $n\ll n_x$, the behaviour is described by $d(n)\propto d_0^{\alpha}$. Since the distance is essentially constant in this regime, the exponent is $\alpha\cong1$, in agreement with the analytical prediction obtained from Eq.~(\ref{eq7}) for short times.

\begin{figure}[t]
\centerline{\includegraphics[width=1.0\linewidth]{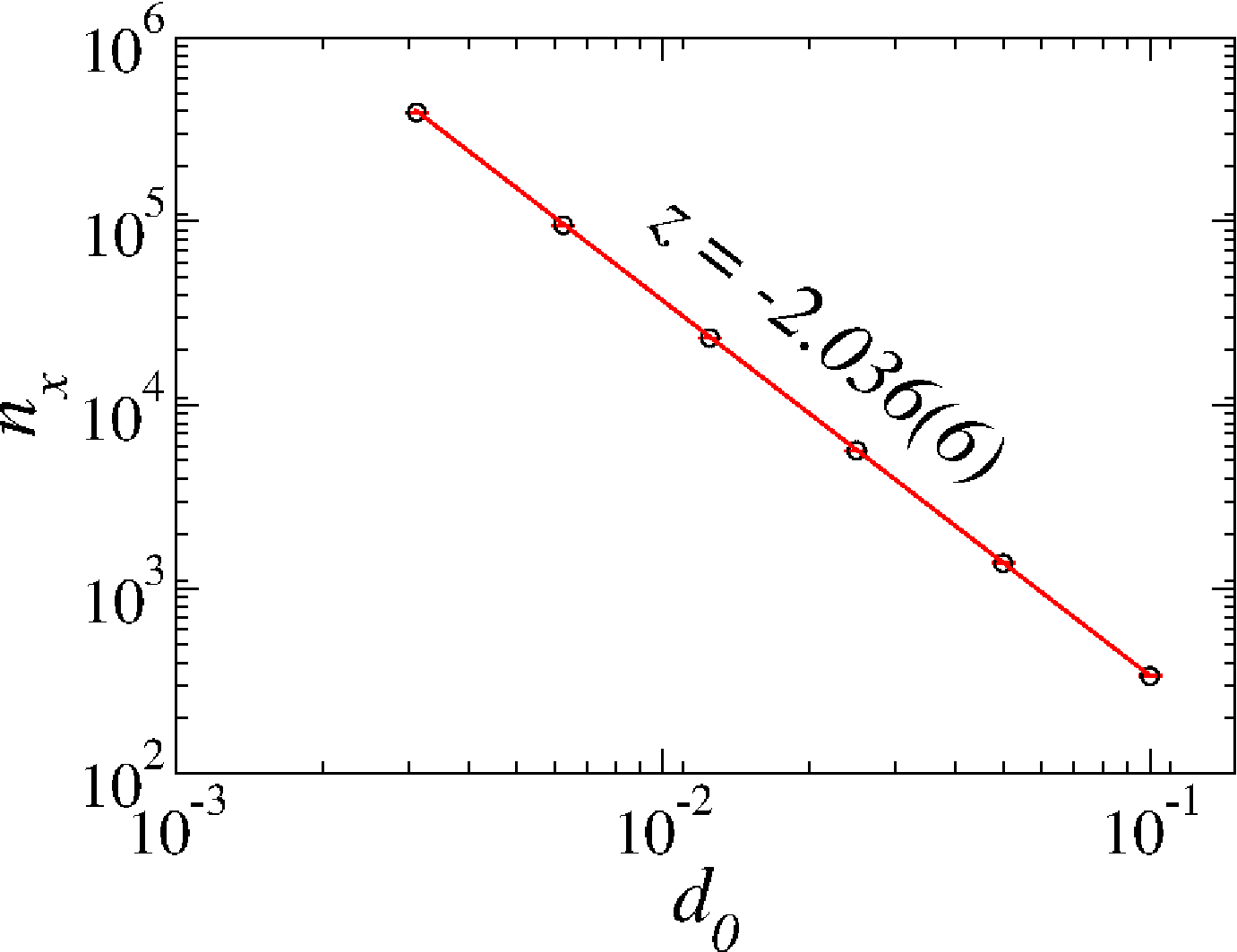}}
\caption{Crossover iteration number $n_x$ as a function of the initial distance $d_0$. A power-law fit yields $z=-2.036(6)$.}
\label{Fig4}
\end{figure}

For sufficiently large $n$, the decay towards the stationary state follows a power law, $d(n)\propto n^{\beta}$, with a numerical slope $\beta=-0.491(3)\approx-1/2$, again in excellent agreement with the analytical prediction. Finally, the crossover iteration number scales as $n_x\propto d_0^{z}$, with $z=-2.036(6)$, as shown in Fig.~\ref{Fig4}.

The universality of the curves shown in Fig.~\ref{Fig3}(a) is confirmed by applying the scaling transformations (i) $d(n)\rightarrow d(n)/d_0^{\alpha}$ and (ii) $n\rightarrow n/d_0^{z}$. As shown in Fig.~\ref{Fig3}(b), these transformations lead to an excellent collapse of all curves onto a single universal curve.

\begin{figure}[t]
\centerline{\includegraphics[width=1.0\linewidth]{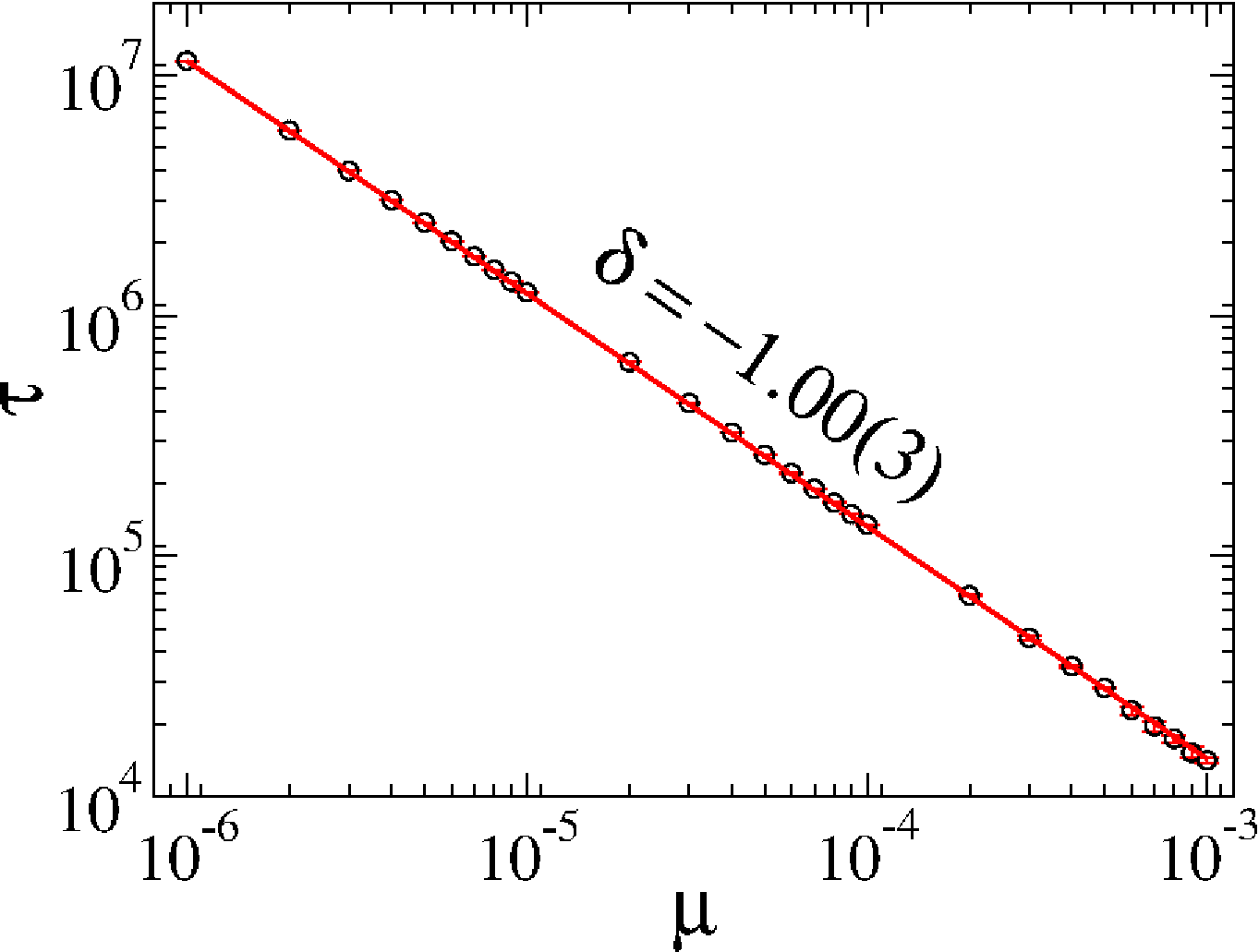}}
\caption{Relaxation time $\tau$ as a function of $\mu$ for the H\'enon map. A power-law fit yields $\delta=-1.00(3)$, in excellent agreement with the theoretical prediction.}
\label{Fig5}
\end{figure}

When the control parameter is chosen such that the dynamics lies close to, but not exactly at, the bifurcation point, the convergence to the fixed point is governed by an exponential decay. In this regime, the characteristic relaxation time scales as $\tau\propto\mu^{\delta}$. Our analytical analysis predicts $\delta=-1$, a result that is fully confirmed by the numerical data shown in Fig.~\ref{Fig5}, where a power-law fit yields $\delta=-1.00(3)$.

The numerical procedure used to obtain the critical exponent $\delta$ is as follows. We consider an ensemble of initial conditions near the fixed point, typically at a distance $d_0=0.01$, and evolve them in time. When the distance of the dynamical variables to the fixed point becomes smaller than a prescribed tolerance -- here taken as $tol<10^{-8}$ -- the corresponding number of iterations is recorded and a new initial condition is selected. This procedure is repeated in order to determine an average convergence time. The control parameter is then updated and the procedure is repeated.

\begin{figure}[t]
\centerline{\includegraphics[width=1.0\linewidth]{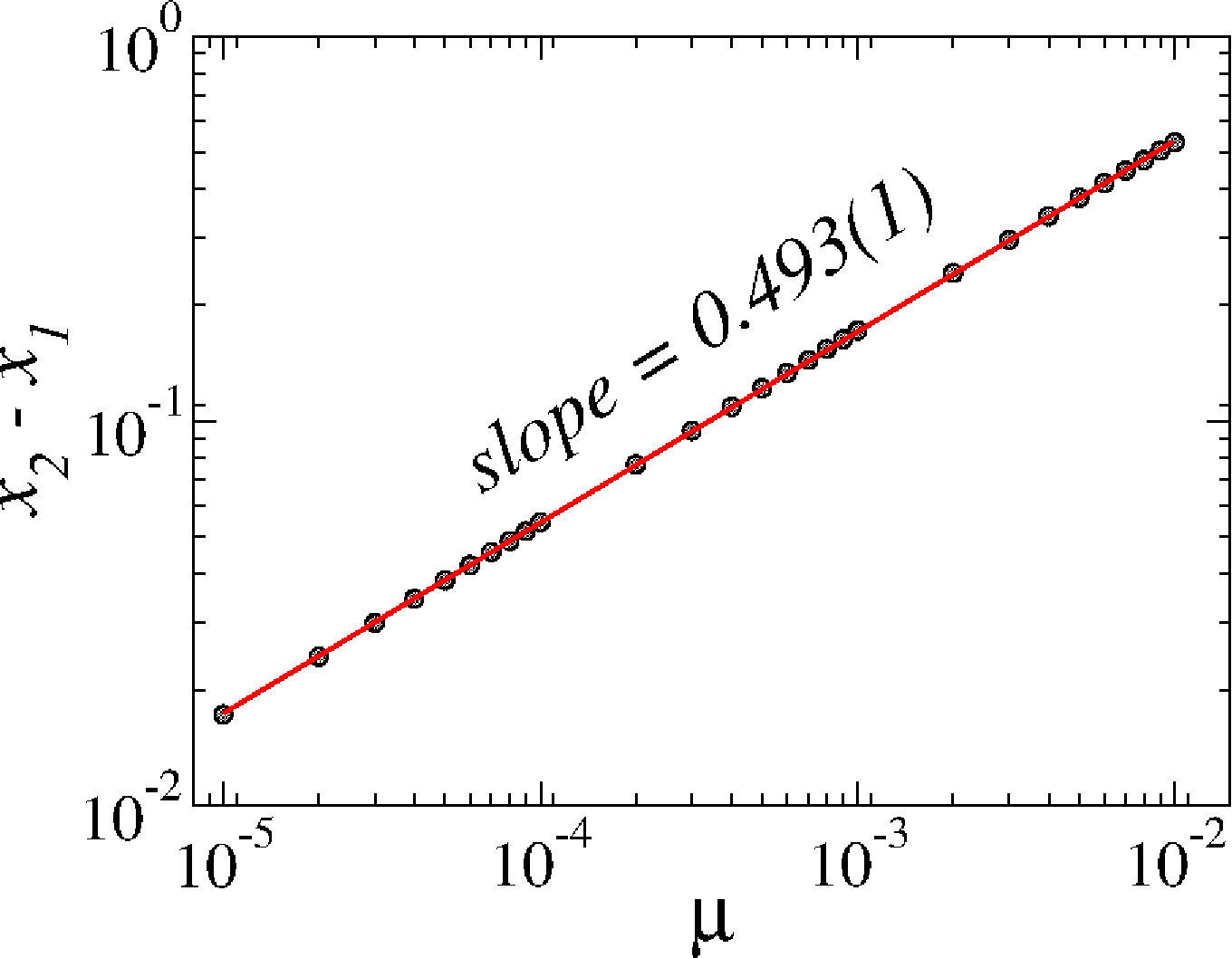}}
\caption{Scaling of the separation between the stable fixed points, $|x_3-x_2|$, as a function of $\mu$ for the H\'enon map. A power-law fit yields a slope of $0.493(1)\cong1/2$, as predicted by the theory. The parameter used was $\mu=a-a_c$ with $a_c=0.36749$.}
\label{Fig6}
\end{figure}

After the bifurcation, the fixed-point analysis indicates that the amplitude of the stable fixed points grows as $\epsilon\sim\sqrt{|\mu|}$. Numerical results for the H\'enon map confirm this behaviour, with the separation between the two stable fixed points scaling as $|x_3-x_2|\propto\sqrt{\mu}$. As shown in Fig.~\ref{Fig6}, a power-law fit yields a slope of $0.493(1)$, which is remarkably close to the theoretical value $1/2$.

\begin{figure}[t]
\centerline{(a)\includegraphics[width=1.0\linewidth]{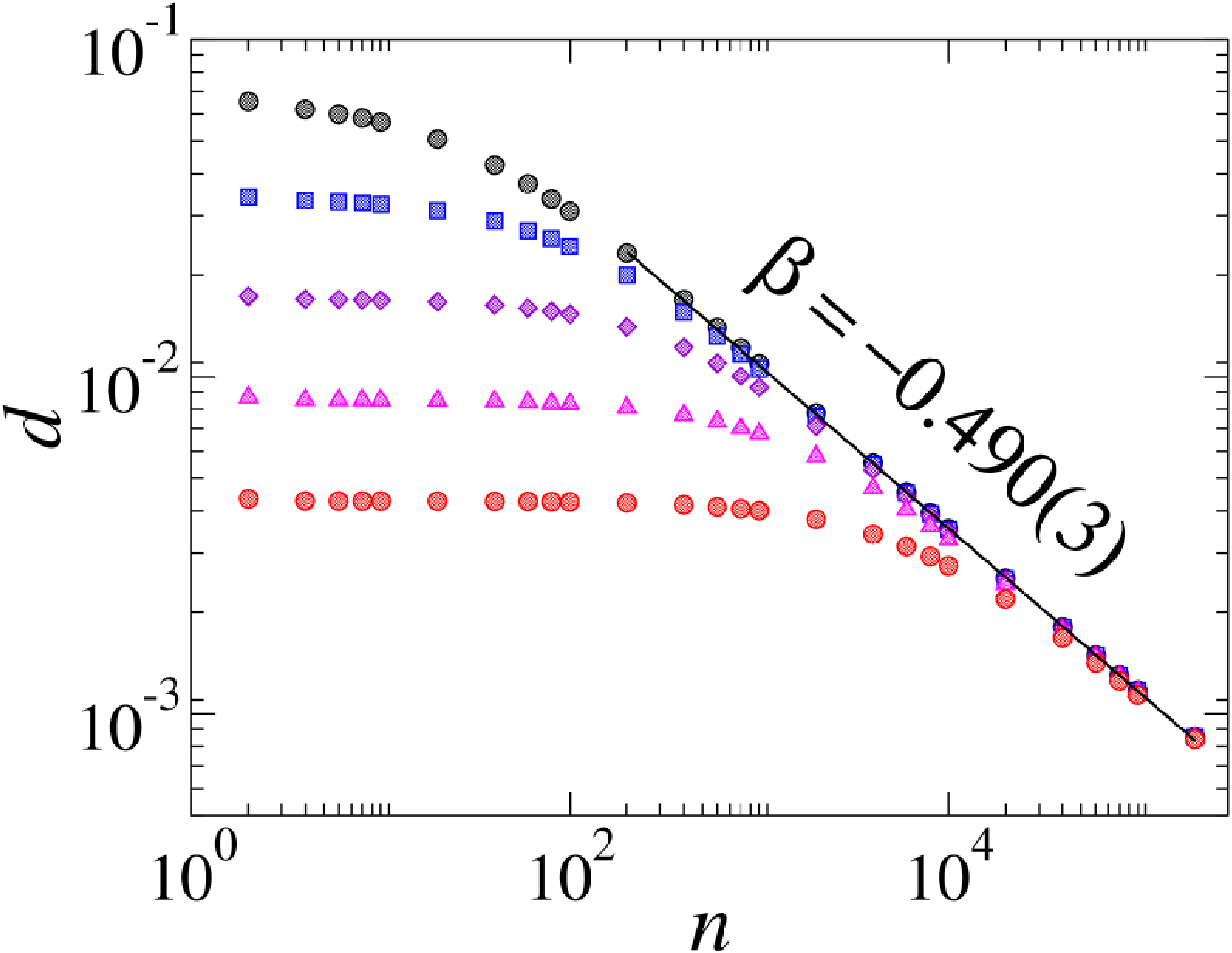}}
\centerline{(b)\includegraphics[width=1.0\linewidth]{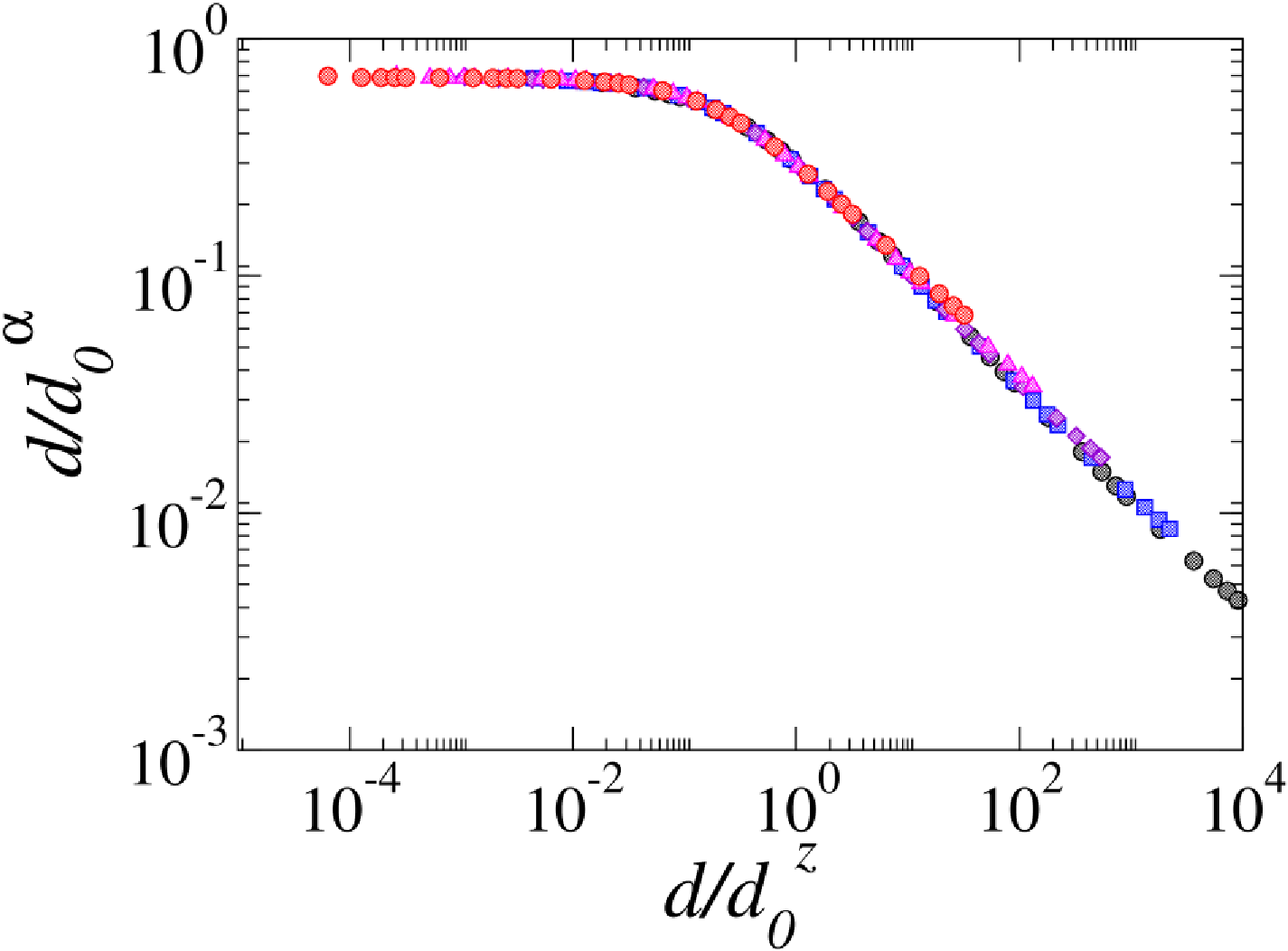}}
\caption{(a) Distance $d(n)$ as a function of the iteration number $n$ for different initial conditions in the vicinity of the fixed point at the period-doubling bifurcation of the Ikeda map. (b) Collapse of the curves shown in panel (a) onto a single universal curve after the appropriate scaling transformations. The control parameters used were: $\theta=0.4$, $\phi=6$, $u_y = 0.6$ and $u_x=0.322191342776$ with initial conditions given by $x_0=1$ and $y_0=0.2$.}
\label{Fig3_ikeda}
\end{figure}

\subsection{The Ikeda map: universality in a second two-dimensional example}
\label{sec:ikeda}

To further support the robustness of the phenomenological description derived in Sec.~\ref{sec2}, we consider a second two-dimensional discrete system exhibiting a period-doubling bifurcation: the Ikeda map \cite{ikeda1979multiple,ikeda1980optical,watanabe1994constants,fm2024mapping}. Originally introduced as a reduced model for light dynamics in an optical cavity, the Ikeda map is commonly written in terms of a complex variable as
\begin{equation}
z_{n+1}=A+B z_n e^{it_n},
\qquad
t_n=\theta-\frac{\phi}{|z_n|^2+1},
\label{eq:ikeda_complex}
\end{equation}
where $z_n=x_n+i y_n$ and $|z_n|^2=x_n^2+y_n^2$. By using Euler's identity $e^{it_n}=\cos(t_n)+i\sin(t_n)$ and separating real and imaginary parts, one obtains a two-dimensional mapping. In the present work, we adopt a dissipative version in which independent dissipation factors are introduced in each component, namely
\begin{eqnarray}
x_{n+1} &=& 1 + u_x\left[x_n\cos(t_n)-y_n\sin(t_n)\right], \nonumber\\
y_{n+1} &=& u_y\left[x_n\sin(t_n)+y_n\cos(t_n)\right],
\label{eq:ikeda_2d}
\end{eqnarray}
with
\begin{equation}
t_n=\theta-\frac{\phi}{x_n^2+y_n^2+1},
\end{equation}
and $u_x,u_y\in[0,1]$. The parameters $u_x$ and $u_y$ control the contraction of phase-space volumes and represent effective losses in the optical cavity. Smaller values correspond to stronger dissipation and typically promote faster relaxation, while values closer to unity reduce losses and allow longer-lived transients and more intricate dynamics.

Our goal here is not to provide an exhaustive bifurcation diagram for the Ikeda map, but rather to verify that, at the first period-doubling bifurcation, the convergence towards the stationary state obeys the same universal scaling laws reported for the H\'enon map in Fig.~\ref{Fig3}. We proceed as follows.For the parameter set $(\theta,\phi,u_y,u_x)=(0.4,6,0.6,0.322191342776)$ and
initial conditions $x_0=1$ and $y_0=0.2$, the map exhibits a period-doubling bifurcation at a critical value of the control parameter $r=r_c$. We compute the corresponding fixed point $(x^\ast,y^\ast)$ at criticality. We then initialise an ensemble of trajectories at different initial distances from the fixed point and monitor the convergence through the Euclidean distance described by Eq. \ref{EDist}
The numerical results for the Ikeda map, shown in Fig.~\ref{Fig3_ikeda}(a), display the same three-regime structure observed previously: (i) an initial plateau in which $d(n)$ remains approximately constant for $n\ll n_x$, (ii) an asymptotic power-law decay for $n\gg n_x$, and (iii) a crossover separating the two regimes at an iteration number $n_x$. In particular, the short-time scaling is consistent with $d(n)\propto d_0^{\alpha}$ with $\alpha\simeq 1$, while the late-time decay follows $d(n)\propto n^{\beta}$ with $\beta\simeq-1/2$, in agreement with the phenomenology derived from Eq.~(\ref{eq7}). Moreover, the crossover iteration scales as $n_x\propto d_0^{z}$ with $z\simeq-2$.

As in the H\'enon case, the universality of the relaxation curves is confirmed by applying the same scaling transformations used to obtain the collapse in Fig.~\ref{Fig3}(b), namely : $d(n)\rightarrow \frac{d(n)}{d_0^{\alpha}}$ and $n\rightarrow \frac{n}{d_0^{z}}$.
After rescaling, all curves collapse onto a single master curve, as shown in Fig.~\ref{Fig3_ikeda}(b). This result provides an independent numerical validation that the period-doubling critical slowing down in two-dimensional maps belongs to the same universality class, even when the microscopic structure of the mapping is substantially different (quadratic polynomial H\'enon versus trigonometric/nonlinear phase coupling in the Ikeda map) and when explicit dissipation is introduced independently in each component.

The Ikeda map results reinforce the central conclusion of the present paper: the local normal form governing a period-doubling bifurcation in two dimensions reduces to the same universal cubic structure in the second iterate, implying that the critical exponents $(\alpha,\beta,z)$ characterising the slowing down at criticality are robust across distinct two-dimensional discrete dynamical systems.

\section{Conclusions}
\label{conc}

In this work, we developed a phenomenological framework to describe the critical slowing down associated with period-doubling bifurcations in discrete dynamical systems. By performing a local expansion around the fixed point and the critical control parameter, we derived a reduced description that captures the convergence towards the stationary state both exactly at, and in the vicinity of, the bifurcation.

At criticality, we identified three critical exponents characterising the short-time behaviour, the asymptotic decay towards the stationary state, and the crossover between these regimes. All three exponents were shown to be universal, as they do not depend on the specific functional form of the underlying mapping. Away from the bifurcation, an additional critical exponent governing the divergence of the relaxation time was obtained, yielding $\delta=-1$ for the period-doubling bifurcation. This result further reinforces the universal nature of critical slowing down in discrete-time systems.

We extended the analysis to two-dimensional mappings and demonstrated that, despite the increased dimensionality and the presence of additional stable directions, the local dynamics near a period-doubling bifurcation is effectively one-dimensional. By projecting the dynamics onto the centre manifold, we showed that the normal form of the second iterate reduces to the same cubic structure encountered in one-dimensional maps. As a consequence, the universal scaling laws derived in one dimension remain valid for higher-dimensional discrete systems undergoing period-doubling bifurcations.

The theoretical predictions were first confirmed numerically using the H\'enon map, for which excellent agreement was found for all critical exponents, scaling relations, and crossover behaviour. To further demonstrate the robustness of the proposed framework, we also analysed the dissipative Ikeda map, a two-dimensional system with a markedly different nonlinear structure and physical origin. Despite these differences, the Ikeda map exhibits the same scaling behaviour at the period-doubling bifurcation, including identical power-law decay, crossover scaling, and data collapse. This additional example provides independent numerical evidence that the critical slowing-down phenomenology is not restricted to polynomial mappings, but extends to systems with trigonometric nonlinearities and phase-dependent coupling.

Earlier numerical evidence for these critical exponents had been reported in the context of a dissipative Fermi--Ulam model~\cite{s2023bifurcations}. The present work places these observations within a unified theoretical framework and shows that they arise naturally from the universal normal form governing period-doubling bifurcations.

Overall, our findings demonstrate that critical slowing down at period-doubling bifurcations can be understood within a robust and unified framework that applies across different system dimensions and dynamical settings. This approach establishes a clear link between bifurcation theory, universality, and scaling behaviour, and can be readily extended to other classes of bifurcations as well as to higher-dimensional discrete dynamical systems.

\section*{Acknowledgments}
E.D.L. acknowledges support from Brazilian agencies CNPq (No. 301318/2019-0, 304398/2023-3) and FAPESP (No. 2019/14038-6 and No. 2021/09519-5). J.P.C.R. thanks FAPESP (2025/20012-0 and 2025/09985-7).


\end{document}